\documentclass[pra,preprint,showpacs,aps]{revtex4}
\usepackage{graphicx}
\begin{document}
 
\title{Magnetic and Transport Properties of $NiFe/SiO_{2}/NiFe$ Trilayers}

\author{J. Gieraltowski and C. Tannous}
\affiliation{ Laboratoire de Magn\'{e}tisme de Bretagne, UMR 6135 CNRS,
Universit\'{e} de Bretagne Occidentale, BP: 809 Brest CEDEX, 29285 FRANCE }

\date{February 11, 2002}

\begin{abstract}

Magnetic coupling between $Ni_{81}Fe_{19}$ double thin films separated by an 
insulating $SiO_{2}$ spacer layer is investigated with structural 
measurements, cross sectional imaging, magnetization measurements (BH loop) 
and transport under the conditions of variable spacer thickness ranging from 
0 to 5000 Angstroms. We investigate several models to explain the coercivity 
variation with spacer thickness as well as the transport measurements.


\end{abstract}

\maketitle

\section*{I. INTRODUCTION}

Mass Storage, Sensing, Spintronics, Quantum computing and other applications 
of magnetic thin-film devices are driving the current interest in 
understanding the nature and extent of magnetic exchange interactions and 
coupling effects arising between layers and nanostructures. While most 
investigations have focused on metal-sandwiched multilayer systems like 
Fe/Cr and Co/Cu where the thickness of the metallic material between the 
ferromagnetic layers determines whether the interlayer coupling will be 
ferromagnetic or antiferromagnetic, this work is focused on ferromagnetic 
layers sandwiching an insulating or semiconducting spacer of variable 
thickness. These systems are of interest in the study of e.g. read heads 
made from multilayered films laminated with insulators (usually 
$Al_{2}O_{3}$) to prevent eddy currents or magnetic tunnel junctions using 
two magnetic layers with different coercivities, separated by a thin 
insulating layer (1 to 2 nm thick). One of the most striking effects of this 
coupling, is the reduction of coercivity of a single layer, attributed to 
reduced domain wall energy, that may alter drastically the device behavior.\\ 

Several possibilities exist and have been studied to describe interactions 
between two ferromagnetic layers across an insulating or semiconducting 
layer \cite{platt} (see Fig. 1). Pinholes generally mediate direct ferromagnetic 
coupling between neighbouring multilayers. They are viewed like shorts 
across the insulating or semiconducting layer. N\'{e}el proposed that conformal 
roughness (orange-peel model) at interfaces can result in ferromagnetic 
coupling for a moderate thickness of spacer material. Recently, an 
uncorrelated roughness model for biquadratic coupling has been proposed as 
well \cite{strijkers}. In N\'{e}el model, roughness features on the surface of the bottom 
ferromagnetic layer propagate through the uppermost layers as they are 
deposited \cite{platt}.\\

Magnetostatic coupling occurs between the roughness features or between 
domain walls in the two magnetic layers. In this case, stray flux fields 
from walls in one film can influence the magnetization reversal process in 
the other. \\

Slonczewski loose-spin model \cite{strijkers} is based on the change in angular momentum 
experienced by spin-polarized electrons tunneling across a spacer barrier 
resulting in a magnetic exchange coupling. The magnitude and sign of the 
coupling oscillates depending on the interfacial barrier properties.

\begin{figure}[htbp]
\begin{center}
\scalebox{0.4}{\includegraphics*{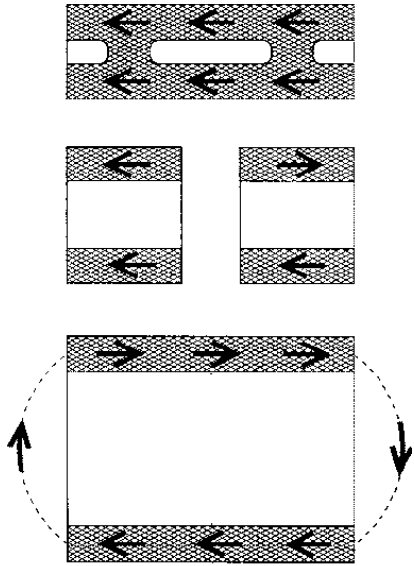}}
\end{center}

\caption{The top drawing shows ferromagnetic coupling arising from pinholes 
existing across the spacer whereas the middle drawing shows exchange coupled 
layers that might be ferro (left) or antiferromagnetic (right). Finally, the 
bottom drawing schematizes magnetostatic coupling between the top and bottom 
layers.}

\end{figure}

This work aim is to collect various structural, layer morphology, magnetic 
(mainly BH loop and coercive field $H_{c}$ measurements) and transport 
evidences in order to discriminate among all these different models.

\section*{II. SAMPLE PREPARATION}

The trilayer samples in this study were grown by RF sputtering onto Corning 
7059 Glass substrates at nominal room temperature. The wafers were mounted 
on a rotating substrate table that passes the wafers over individual 
sputtering guns. The sputtering gas used is $3.10^{-3}$ mbar of Ar in a 
background pressure of $1.10^{-6}$ mbar and an RF power density of 
1W/cm$^{2}$. The base magnetic layer is 480 \AA \hspace{1mm} of Permalloy 
($Ni_{81}Fe_{19}$ alloy denoted as Py) with $SiO_{2}$ as the spacer 
material, and another top layer of Py 480 \AA \hspace{1mm} thick. Each magnetic film was 
sputtered from a composite target. The $SiO_{2}$ was also RF sputtered from 
an oxide target at a power density of 1W/cm$^{2}$. The thickness of the Py 
layers was monitored precisely and kept constant. Its value was carefully 
chosen in order to minimise H$_{c}$ for this type of structure (In Fig.2 the 
morphology of the trilayer is shown with a special High Resolution TEM 
technique). The magnetization of all the Py films was planar in all the 
samples studied. 

\begin{figure}[htbp]
\begin{center}
\scalebox{0.4}{\includegraphics*{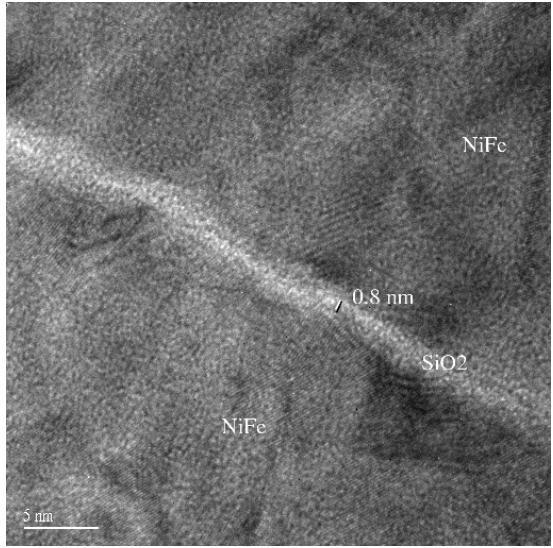}}
\end{center}
\caption{ Cross sectional high resolution transmission electron microscopy 
(HRTEM) image of the trilayer structure showing clearly the $SiO_{2}$ 
sandwich layer 0.8 nm thick (in this case) between the Py layers.}
\end{figure}

Clearly, a small undulation is visible in the TEM image (with an average 
amplitude h $ \approx $ 1 nm and wavelengh $\lambda  \approx $ 30 nm) 
indicating that the roughness of the $SiO_{2}$ layer at the two interfaces 
with the Py layers is conformal and might play a role in the physical 
properties as discussed later.

\begin{figure}[htbp]
\begin{center}
\scalebox{0.9}{\includegraphics*{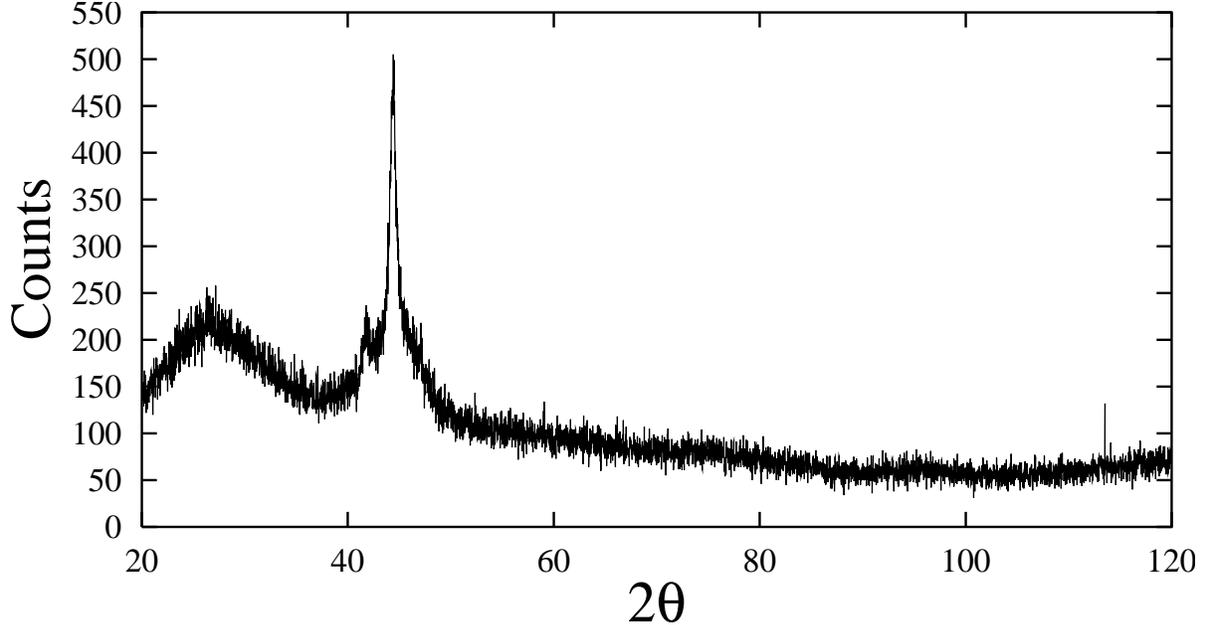}}
\end{center}

\caption{ Typical X-Ray diagram of the structure analysed (2 $\theta $ in 
degrees). The broad features is an indicator of an overwhelming presence of 
an amorphous phase whereas the sharp peak corresponds to the scattering by a 
microcrystallite of Py.}

\end{figure}

The amorphous nature of the structure is demonstrated by the X-Ray diagram 
shown in Fig.3. The sharp feature appearing in the diagram corresponds to 
the presence of microcrystallites of Py type with a textured growth along 
(111) direction.

\section*{III. HYSTERESIS LOOP MEASUREMENT}

The trilayers showed marked anisotropic magnetization behavior in hysteresis 
BH loop measurements. On the average, the anisotropy field observed in the 
the hard axis configuration of the BH loop was about 4 Oe. That is not the 
case of the easy axis coercivity as described in the next section. We 
believe that our samples possess a uniaxial anisotropy character as shown by 
Platt et al. \cite{platt} whose high-field torque measurements confirm the presence 
of a uniaxial anisotropy, under similar conditions.\\

Several sets of BH loop measurements were performed in addition to 
understand and assess the true nature of the coupling between the layers. We 
performed measurements on raw films, uncoupled layers, saturated layers and 
finally fully coupled layers. 

\section*{IV. COERCIVITY MEASUREMENTS}

Magnetic coupling between double thin films separated by the insulating 
$SiO_{2}$ spacer layer is investigated with magnetization measurements (BH 
loop) under the conditions of variable $SiO_{2}$ spacer thickness ranging 
from 0 to $\sim $5000 \AA. \\

The easy axis coercive field H$_{c}$ displayed in Fig. 4 decreases, then 
shows a fine structure in the form of damped oscillations to increase again 
with the thickness of the $SiO_{2}$ spacer layer. The drop of the coercivity 
H$_{c}$ compared to the free Py layer may attain a factor of 15 under 
certain conditions. For larger values of $SiO_{2}$ thickness, the 
magnetization reversal becomes sharper. In addition, when the spacer layer 
is thicker than 100 \AA, separate reversal of the magnetic layers is observed 
in the hysteresis loop.

\begin{figure}[htbp]
\begin{center}
\scalebox{0.9}{\includegraphics*{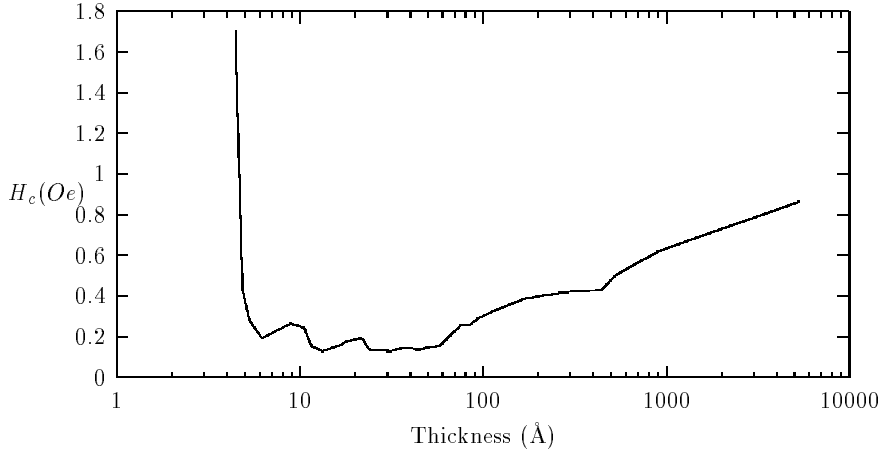}}
\end{center}

\caption{Easy axis coercivity variation with the thickness of the spacer 
$SiO_{2}$ layer.}
\end{figure}

\section*{V. TRANSPORT MEASUREMENTS}

Magnetoresistive devices have transport properties that depend on the 
relative orientation of magnetic moments in neighboring grains or thin 
films. Application of an external magnetic field along different axes (easy 
and hard axes of Py) and with respect to the injected current direction 
\textbf{J} results in a specific dependence of scattering on spacer 
thickness. The key to understanding of the scattering is based on domain 
wall orientation in the layers \cite{viret}. \\

The in-layer magnetic domain wall  structure is the source of scattering for
 the injected electrical current in 
the structure. Fig. 5 displays, respectively, the easy axis (EA) coercive 
field H$_{c}$ with a damped oscillating fine structure (for an $SiO_{2}$ 
thickness 0.6 to $ \sim $3 nm) and the corresponding magnetoresistance ratio 
MR values defined as $\mbox{MR} = \frac{R_{S}- R_{0}}{R_{0}} = \frac{\Delta R}{R_{0}}$ 
where $R_{S}$ et $R_{0}$ are measured with and without the saturation 
magnetic field. Magneto-transport data were taken with a high resolution 
four-point probe with the current \textbf{J} flowing either parallel to the 
domain wall, i.e. parallel to the easy axis (EA) or perpendicular to the 
domain wall, i.e. parallel to the hard axis (HA). The magnetic field 
\textbf{B} is applied either parallel to the current \textbf{J} direction 
(\textbf{B} $\parallel$ \textbf{ J}) or perpendicular to \textbf{J} (\textbf{B} $\perp$ 
\textbf{J}). The coupling between the layers alters the domain wall 
configuration and thus modulates the scattering affecting the 
magnetoresistance. Different orientation of the domain walls for different 
thickness has been studied extensively in the past \cite{feldtkeller} and the transport 
data correlate well with H$_{c}$ fine structure as discussed next.

\begin{figure}[htbp]
\begin{center}
\scalebox{0.9}{\includegraphics*{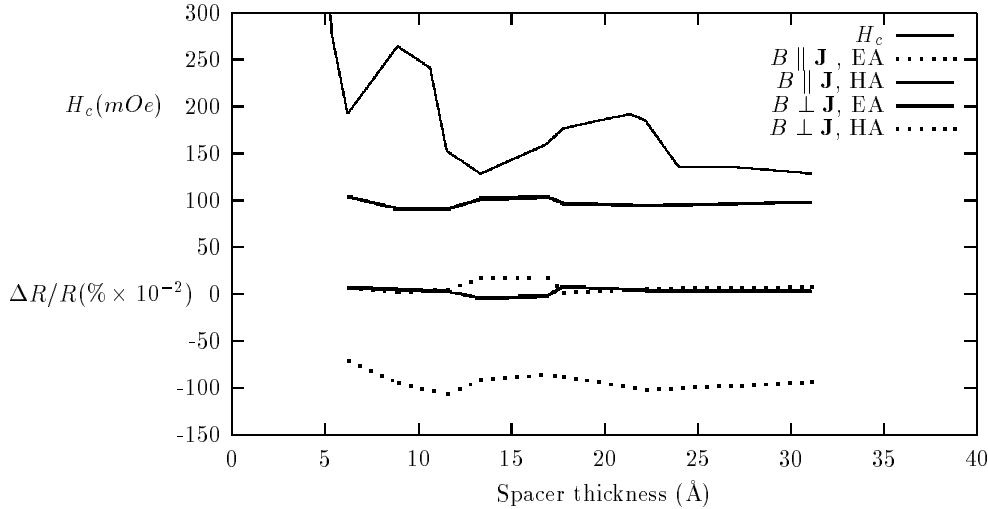}}
\end{center}

\caption{Easy axis coercivity (top curve) and magnetoresistance ratio (four lower curves) for various orientations of the magnetic field with respect to current and anisotropy axis versus the thickness of the $SiO_{2}$ spacer layer.}
\end{figure}

\section*{VI. DISCUSSION AND CONCLUSIONS}

Our study reveals that the EA coercivity field is a powerful probe to test 
magnetization reversal processes in magnetically ordered and artificially 
structured magnetic materials. For any given material H$_{c}$ depends on 
coupling reflecting the average properties of the sample as well as local 
interactions such as wall-wall interaction or magnetization anisotropy 
dispersion (ripples), in addition to some intrinsic quantities (such as the 
effective anisotropy and saturation magnetization). This paves the way to 
manufacture ultra-soft artificial magnetic nanostructures, e.g. our 
structure was so soft (in comparison with a single layer of same thickness) 
that the fine structure in H$_{c}$ may be viewed as some kind of magnetic 
imaging of wall topography in coupled thin films. \\

The observed decrease in H$_{c}$ is more pronounced for the EA field 
direction, where the magnetization process is determined by domain wall 
motion. The internal flux closure of the N\'{e}el-wall like structures leads to 
a reduced interaction in the film and between walls in coupled layers, 
resulting in a drop of H$_{c}$. In addition, we believe that wall character 
changes are driven by coupling in the following way. If the magnetization is 
reversed by wall motion, the effect of the coupling will be that both Py 
film magnetizations are reversed simultaneously. This means that a wall in 
one film is always close to another in the other film. Because of the dipolar 
magnetic moment of the N\'{e}el-walls there will be a magnetostatic interaction 
between the walls (in addition to an eventual coupling in the case 
of loose spins at the interface or pinhole presence). Following \cite{feldtkeller}, the 
resulting configurations are shown in Fig. 6.

\begin{figure}[htbp]
\begin{center}
\scalebox{0.3}{\includegraphics*{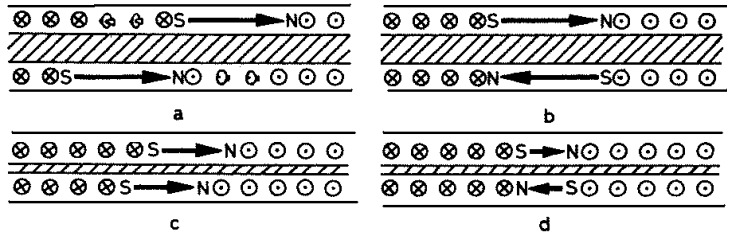}}
\end{center}

\caption{Domain wall configurations in a multilayer magnetic film that 
contains a thin nonmagnetic spacer between two ferromagnetic films of same 
thickness. In c and d the spacer is thinner than a and b.}

\end{figure}

Two principally different configurations must be considered, one with the 
magnetization of both walls parallel (Fig. 6a and c), and another with 
anti-parallel magnetizations. When the coupling is weak, the arrangement 
will be governed by magnetostatic interaction yielding the configuration 
shown in Fig. 6a and b. For a stronger coupling (due to a thinner 
intermediate layer (or with loose spins at interfaces or pinholes), the 
configurations shown in Fig. 6c and d must be expected. The configuration of 
Fig. 6c and d will have a wall energy larger than those of Fig. 6a and b. In 
addition, specific energy conditions in the film (generally of magnetostatic 
origin), favor a local ferro or antiferromagnetic-like coupling (see Fig. 
5). We estimate following \cite{bobo} and by looking at Fig.2 that 
orange-peel coupling, induced by the roughness of the layers is 
negligible in our case. \\

In conclusion, our work illustrates the detailed study of the nature of the 
coupling between two magnetic layers separated by an insulator spacer of 
variable thickness.\\

{\bf Acknowledgements}:
The authors thank J. Ostorero, F. Michaud and X. Castel for the X-Ray 
measurements and E. Snoeck for the cross section TEM microscopy.

\end{document}